\def\S {\sum_{\l=1}^2}                            
\def\SP {\sum_{\l'=1}^2}                            
\def\aatc {\Big({2 a \tau \over c}\Big)}            
\def\atc {\Big({a \tau \over c}\Big)}               
\def\C {{\hbar \omega^3 \over 2 \pi^2 c^3} d\omega} 
\def\hzp {h_{zp}(\omega)}     
\def\Hzp {H_{zp}(\omega)}      
\def\Hzpsq {H^2_{zp}(\omega)}  
  
\def\l {\lambda}               
\def\w {\omega}                
\def\b {\beta}                 
\def\g {\gamma}               
\def\t {\tau}                  
\def\a {\alpha}                
\def\ad {\alpha^{\dagger}}    
\def\kh {{\hat k}}             
\def\eh {{\hat \epsilon}}      
\def\xh {{\hat x}}             
\def\yh {{\hat y}}             
\def\zh {{\hat z}}           
\def\kbf {{\bf k}}             
\def\kl {(\kbf,\ \l)}         
\def\klp {(\kbf',\ \l')}      
\def\ZL {\langle 0 |}          
\def\ZR {| 0 \rangle }      
\def\rbf {{\bf r}}             
\def\gt {\gamma_\tau}          
\def\Tarr {\buildrel\leftrightarrow\over {\bf T}} 

\font\bbf=cmbx12
\parindent 0 truein
{\bbf Quantum Vacuum and Inertial Reaction in Nonrelativistic QED}

\vskip 10 pt
{\bf Hiroki Sunahata}$^1$, {\bf Alfonso Rueda}$^2$ \footnote{$^\star$}{Send correspondence to this author.}
{\bf and Bernard Haisch}$^3$

\vskip 10 pt
$^1$ Dept. Physics and Astronomy, California State Univ., Long Beach, CA 90840

$^2$ Dept. Electrical Engineering, California State Univ., Long Beach, CA 90840 email: arueda@csulb.edu

$^3$ Calphysics Inst., 519 Cringle Drive, Redwood Shores, CA 94065 email: astro@calphysics.org

\vskip 10 pt
{\bf Abstract}

The possible connection between the electromagnetic zero-point field (ZPF) and the inertia reaction force was first
pointed out by Haisch, Rueda, and Puthoff ({\it Phys.~Rev.~A}, {\bf 49},~678,~1994), and then 
by Rueda and Haisch following a totally different and more satisfactory approach ({\it Found.~of~Phys.}, {\bf
28},~1057,~1998; {\it Phys. Letters A}, {\bf 240}, 115, 1998; {\it Annalen der Physik} {\bf 10}(5), 393, 2001). In the
present paper, the approach taken by Rueda and Haisch will be followed, but the analysis will be done within a
formulation that uses nonrelativistic quantum electrodynamics with the creation and annihilation operators rather than
the approach of Rueda and Haisch using stochastic electrodynamics.  We analyze the interaction between the zero-point
field  and an object under hyperbolic motion (constant proper acceleration), and find that there arises a reaction force
which is proportional in magnitude, and opposite in direction, to the acceleration. This is suggestive of what we know
as inertia. We also point out that the equivalence principle -- that inertial mass and gravitational mass have the same
values -- follows naturally using this approach. Inertial mass and gravitational mass are not merely equal, they are the
identical thing viewed from two complementary perspectives ({\it Annalen der Physik} {\bf 14}(8), 479, 2005). In the
first case an object accelerating through the electromagnetic zero-point field experiences resistance from the field. In
the case of an object held fixed in a gravitational field, the electromagnetic zero-point field propagates on curved
geodesics, in effect accelerating with respect to the fixed object, thereby generating weight. Hence, the equivalence
principle does not need to be independently postulated.

{\bf Keywords:} quantum vacuum, mass, zero-point field, inertia, gravitation, stochastic electrodynamics

\parindent 0.4 truein

\vskip 10 pt
\noindent
{\bf 1. Introduction}

\vskip 10 pt
\noindent
The so-called zero-point field (ZPF) is a random electromagnetic field that exists
even at the temperature of absolute zero. The existence of this field
first came to be known through the study of the blackbody radiation spectrum
early in the twentieth century [1], and became gradually better 
understood with the advance of quantum theory.
Moreover, the developments of Stochastic Electrodynamics (SED) in the last
decades of the twentieth century have expanded its boundary and found new applications. Rueda,
Haisch and Puthoff proposed that the origin of inertia could be
explained, in part, as due to the interaction between an
accelerated object and the zero-point-vacuum-fields. (Only the electromagnetic contribution has been studied so far from
this viewpoint.) In their first approach [2], the Lorentz force that the electromagnetic zero-point 
field (ZPF) exerts upon an 
accelerating harmonic oscillator was calculated, and in the
second by Rueda and Haisch [3], a more general method was taken
by analyzing the
zero-point-field Poynting vector that an extended accelerating object 
sweeps through. In this paper, the second
method will be followed using a {\it Quantum Electrodynamics} (QED) 
approach for all the averaging calculations. We use the low energy version of QED, also called nonrelativistic
QED [4]. It will be shown that the same results reported in Rueda and Haisch [3] are obtained
with QED as well: 
There is a contribution to the inertia reaction force coming from the 
electromagnetic quantum vacuum. Other contributions are naturally 
expected from the other quantum vacuum fields manifested in the so called 
physical vacuum when it is taken as a medium. We
call this the {\it quantum vacuum inertia hypothesis.}

The general objective of this research program is and has been to elucidate the mechanism behind the appearance of an
inertial reaction force when a macroscopic body is being accelerated by an external agent. There are several
contributing factors that independently contribute but so far only the electromagnetic vacuum contribution has been
preliminarily explored. This contribution, although relatively minor, should display some common features that we
expect will help uncover the mechanism behind the contributions from other vacua, namely those of the strong and the
weak interactions. For example, it is ordinarily expected that the gluonic fields of the strong interaction should
give a major contribution to the mass of macroscopic bodies. Though the {\it how} of the corresponding part of the
inertial reaction due to the gluonic fields is not well understood, it is expected that elucidating the mechanism
for the electromagnetic field contribution will help in finding the corresponding mechanism for the more involved
gluonic field.

There are however some common misconceptions as to the manner in which the inertial reaction force appears. There is,
for example, the idea that the vacuum inertia hypothesis implies some ram-pressure-like action by the vacuum fields on
the accelerated body. The analysis of [3] does not support such views.

In this research it is found that the electromagnetic vacuum exerts a peculiar opposition to any change in the change
of a body comoving inertial frame. This opposition is in a direction opposite to that of the acceleration and of a
magnitude proportional to that acceleration. And, very importantly, it is instantaneously applied to all points within
the body. So this means that the opposition is not coming from the outside as a kind of wind or ram-pressure. It is
easy to see that the shape of a body or its relative orientation with regard to the acceleration vector makes no
difference whatsoever.

In section 2, we review the similarities and differences between the SED and 
QED formulations. In section 3, using a QED formulation,
the inertia reaction force is obtained. 
We find that there exists an interaction between the 
ZPF and an accelerating object, and that the reaction force is proportional 
to the acceleration in magnitude, but opposite to the acceleration in direction. The vacuum seems to react against 
accelerating objects in a manner reminiscent of what is 
understood as inertia.

\vskip 10 pt
\noindent
{\bf 2. Comparison between the Quantum and the Stochastic Formalism}

\vskip 10 pt
\noindent
We proceed from a comparison between the quantum formalism of QED and the 
classical formalism of SED. 
The {\it classical} electromagnetic zero-point radiation can be written, 
as a superposition of plane waves [5]:

$$\eqalignno {
{\bf E}({\bf r},t) & = 
\S \int
d^3 k  \ \hat{\epsilon} \kl \hzp 
\cos \lbrack {\bf k \cdot r} - \w t - \theta \kl \rbrack \ , 
& (1) \cr
{\bf B}({\bf r},t) & = 
\S \int
d^3 k \left( {\hat k} \times {\hat \epsilon}\right) \hzp
\cos \lbrack {\bf k \cdot r} - \w t - \theta \kl \rbrack \ .
& (2) \cr } $$

\noindent
Here, the zero-point radiation is expressed in expansion of plane waves 
and as a sum over two polarization states 
$\hat{\epsilon}({\bf k}, \lambda)$, which is a function of the 
propagation vector ${\bf k}$ and the polarization index $\lambda=1,2$.  

From now on and in the above, the polarization components 
${\hat \epsilon}_i({\bf k}, \lambda)$
are to be understood as scalars. They are the projections of the 
polarization unit vectors ${\hat \epsilon}({\bf k},\lambda)$ onto the $i$-axis:
${\hat \epsilon}_i({\bf k}, \lambda)  = {\hat \epsilon} \cdot {\hat x}_i$.
We also use the same notational convention with the ${\hat k}$ unit vector, i.e., 
${\hat k}_x = {\hat k} \cdot {\hat x}$.

In the expressions (1) and (2), the random phase 
$\theta({\bf k}, \lambda)$  
is introduced, following Planck [6],
and Einstein and Hopf 
[7], to generate the random, fluctuating nature of the radiation. 
This $\theta({\bf k}, \lambda)$ is a random variable distributed 
uniformly in the interval  
$(0, 2\pi)$ and independently for each wave vector ${\bf k}$ and the 
polarization index $\lambda$. 
Also the spectral function $\hzp$ is introduced to set the magnitude of 
the zero-point radiation, which is found in terms of the Planck's 
constant $h$ as $h_{zp}^2(\w) = {\hbar \omega}/{2\pi^2}$ [5].
Planck's constant enters the theory at this point only as a scale factor 
to attain correspondence between 
zero temperature random radiation of 
(classical) stochastic electrodynamics and
the vacuum zero point field of 
quantum electrodynamics. 

The QED formulation of the zero point fields are 
also expressed by the expansion in plane waves as [8,9]

$$\eqalignno {
{\bf\overline E}({\bf r},t) & = 
\S \int
d^3 k  \ \hat{\epsilon} \kl \Hzp 
\lbrack \a \kl \exp(i \Theta)+\ad \kl \exp(-i \Theta)
\rbrack \ ,
& (3) \cr
{\bf\overline B}({\bf r},t) & = 
\S \int
d^3 k \left( {\hat k} \times {\hat \epsilon}\right) \Hzp
\lbrack \a \kl \exp(i \Theta)+\ad \kl \exp(-i \Theta)
\rbrack \ ,
& (4) \cr } $$

\noindent
where $\Theta = {\bf k} \cdot {\bf r} - \omega t$.

We notice here that the cosine functions used in the SED formulation are now 
replaced by the exponential functions and the quantum operators 
$\alpha ({\bf k},\lambda)$ and $\alpha^\dagger ({\bf k},\lambda)$. 
These annihilation and creation operators have the expectation values:

$$\eqalignno {
\ZL \a \kl \ad \klp \ZR & =
\delta_{\l,\l'} \delta^3 (\kbf-\kbf') \ ,
& (5a) \cr
\ZL \ad \kl \ad \klp \ZR & =
\ZL \a \kl \a \klp \ZR =
\ZL \ad \kl \a \klp \ZR = 0 \ .
& (5b) \cr} $$

\noindent
The overline on $E$ and $B$ in Eq. (3) and (4)
indicates that these fields are now given as operators, and 
the spectral function in QED is $H_{zp}^2(\w)  = {\hbar \omega}/{4\pi^2}$ [10], which differs, by a small factor, from
the corresponding SED spectral function we used in previous papers [3].

It is well known that 
only in a certain limited set of cases do SED and QED give the same 
results [11].  
Almost forty years ago, T.~H. Boyer [8] presented a detailed 
comparison between these two theories for the case of {\it free electromagnetic 
fields} and for {\it dipole oscillator systems}. In his comparison, it 
was found that if the QED operators are symmetrized 
(written in symmetric order), then the stochastic averaging of SED and 
the quantum averaging of QED yield exactly the same results.

This last point is of much importance for our developments. 
The SED stochastic averaging over the random phases yields for the 
electric field autocorrelation function, at two different space-time 
locations $({\bf r}_1, t_1)$ and $({\bf r}_2, t_2)$, an expression of the form

$$
\langle E_i (\rbf_1,t_1) E_j(\rbf_2,t_2) \rangle =
\int d^3k ( \delta_{ij}-{\hat k}_i {\hat k}_j ) {\hbar \w \over 4\pi^2}
\cos \lbrack \kbf \cdot (\rbf_1  - \rbf_2) - \w(t_1-t_2) \rbrack \ ,
\eqno(6)$$

\noindent
where the subindices $i$ and $j$ stand for any two different cartesian space 
directions, $i,~j = x,~y,~z$ and the $\left< \cdots \right>$ parentheses mean 
a stochastic averaging.
On the other hand, if we do a simple quantum averaging over the vacuum 
field we get

$$
\ZL {\overline E}_i(\rbf_1,t_1) {\overline E}_j(\rbf_2,t_2) \ZR = 
\int d^3k ( \delta_{ij}-{\hat k}_i {\hat k}_j ) {\hbar \w \over 4\pi^2}
\exp \lbrack i\kbf \cdot (\rbf_1  - \rbf_2) - i\w(t_1-t_2) \rbrack \ ,
\eqno(7)$$

\noindent
which clearly shows that $\left< \cdots \right>$ of Eq. (6) and $\ZL\cdots\ZR$ of Eq. (7) are {\it not the same}.
This is of course not surprising. 
The stochastic averaging of Eq. (6) involves 
averaging over the random phases in a manner thoroughly described in 
Refs. [2,3,8,11,12,13].
On the other hand, the averaging described in Eq. (7) is the 
standard quantum averaging where the $\overline{E}_i$ and 
$\overline{E}_j$ fields are taken as operators in the Heisenberg 
picture. Nevertheless, if instead of writing the operator fields as in Eq. (7), 
we write them in terms of a symmetrized expression, then we have that

$$\eqalignno {
\langle E_i(\rbf_1,t_1) E_j(\rbf_2,t_2)\rangle & = 
{1 \over 2}
\Big[ \ZL {\overline E}_i(\rbf_1,t_1) {\overline E}_j(\rbf_2,t_2) \ZR +
 \ZL {\overline E}_j(\rbf_2,t_2) {\overline E}_i(\rbf_1,t_1) \ZR \Big]
& \cr
& = 
\Big\langle 0 \Big| 
{ {\overline E}_i(\rbf_1,t_1) {\overline E}_j(\rbf_2,t_2) +
  {\overline E}_j(\rbf_2,t_2) {\overline E}_i(\rbf_1,t_1) \over 2 }
\Big| 0
\Big\rangle \ .
& (8) \cr }$$

In Refs. [3,10], correlations of the form
$\ZL\overline{E}_i({\bf r}_1,t_1) \overline{B}_j({\bf }_2,t_2)\ZR$
were calculated in an effort to evaluate 
$\ZL{\overline{\bf E}\times\overline{\bf B}}\ZR$.
In this case as well, if the quantum operators are properly symmetrized, 
the stochastic averaging of SED and the quantum averaging of QED give 
identical results. See Appendix A for 
more details.

\vskip 10 pt
\noindent
{\bf 3. Origin of the Electromagnetic Vacuum Contribution to the Inertia Reaction Force}

\vskip 10 pt
\noindent
Let us consider an object to be uniformly accelerated by a force applied to it 
by an external agent and such that the object moves rectilinearly along 
the $x$-axis with constant proper acceleration ${\bf a} = \hat{x}a$. 
We need only look at the coordinates of the center of mass and for most 
purposes view the object as point-like.
The object then performs so-called hyperbolic motion [14,15]. 
Assume the body was instantaneously at rest at time $t_* = 0$ in an 
inertial frame $I_*$ that we call the laboratory frame. 
Consider a non-inertial frame $S$ such that its $x$-axis coincides 
with that of $I_*$ and let the body be located at coordinates $(c^2/a, 0, 
0)$ in $S$ at all times. 
So this point of $S$ performs hyperbolic motion. The acceleration 
of the body point in $I_*$ is ${\bf a}_*=\gamma_{\tau}^{-3}{\bf a}$ 
at body proper time $\tau$. 
We take $S$ as a rigid frame and therefore only neighboring points 
of $S$ around the body are found to have the same acceleration. 
The frame $S$ we call the Rindler frame. Consider also an infinite 
collection of inertial frames $\{I_\tau\}$ such that at body proper time $\tau$, 
the body is located at point $(c^2/a, 0, 0)$ of $I_\tau$. The $I_\tau$ frames 
have all axes parallel to those of $I_*$ and their $x$-axes coincide with 
that of $I_*$. We set the proper time $\tau$ such that at $\tau = 0$ 
the corresponding $I_\tau$ coincides with $I_*$. So clearly 
$I_{\tau=0} = I_*$. If this is so then the hyperbolic motion guarantees that

$$\eqalignno {
x_* & = {c^2 \over a} \cosh \atc \ , & (9) \cr
t_* & = {c \over a}   \sinh \atc \ , & (10) \cr
\b_{\t} & = {u_x(\t) \over c} = \tanh \atc \ ,& (11) \cr
\g & = (1 - \b^2)^{-1/2} = \cosh \atc \ .
& (12) }$$

In what follows, we reproduce a brief sketch of the derivation of the 
electromagnetic contribution to the inertia reaction force in QED 
formulation [3,10,12]. 
It will be shown, as indicated in Eq. (34), that
the final averaged results turn out to be the same for both developments (SED and QED).

The QED formulation of the zero-point electric and magnetic fields are 
given in Eq. (3) and Eq. (4).
We now Lorentz transform these fields from the laboratory frame $I_*$ 
into an instantaneously comoving frame $I_\tau$ to calculate 
the EM zero-point field vectors ${\bf E}_{zp}$ and ${\bf B}_{zp}$ of 
$I_*$ but as represented in $I_\tau$. 

$$\eqalignno {
{\bf E}_\t^{zp} (0,\t)
& = \sum _{\l=1}^2 \int d^3k \ \Hzp
\Big\{ \xh \eh_x + \yh \cosh \atc \Big[\eh_y - \tanh \atc (\kh \times \eh)_z \Big]
& \cr 
& + \zh \cosh \atc \Big[ \eh_z + \tanh \atc (\kh \times \eh)_y \Big] \Big\}
\Big\{\a \kl e^{i \Theta} + \ad \kl e^{-i \Theta}
\Big\} ,
& (13) \cr }
$$

$$\eqalignno {
{\bf B}_\t^{zp} (0,\t)
& = \sum _{\l=1}^2 \int d^3k \ \Hzp
\Big\{ \xh (\kh \times \eh)_x + \yh \cosh \atc \Big[(\kh \times \eh)_y- \tanh \atc \eh_z
\Big] & \cr 
& + \zh \cosh \atc \Big[ (\kh \times \eh)_z + \tanh \atc \eh_y \Big] \Big\}
\Big\{\a \kl e^{i \Theta} + \ad \kl e^{-i \Theta}
\Big\} ,
& (14) \cr }
$$

\noindent
where $\Theta$ is given by

$$\Theta = \kh_x {c^2 \over a} \cosh \atc - \w {c \over a} \sinh \atc .
\eqno(15).
$$

\noindent
Here, unlike the classical random variable cases, the order of the quantum 
operator affects the results in the formulation, which is a major 
difference between the previous SED [3] and this present QED treatment.

We assume these fields as seen in $I_\tau$ to also correspond to the fields as 
{\it instantaneously} seen in $S$ at proper time $\t$. Though the fields at the object point 
in $S$ and in the corresponding point of the co-moving frame $I_\tau$ 
that instantaneously coincides with the object point are the same, this 
does not mean that detectors in $S$ and in $I_\tau$ will experience the same 
radiation-field time evolution. 
A detector at rest in $I_\tau$ 
and the same detector at rest in $S$ do not experience timewise the same 
effect. The two fields 
are the same at a given space-time point; however, the time evolution 
and space distribution of the field in $S$ and those of the field in  
$I_\tau$ are not the same.  

We consider next the ZPF radiation background of $I_*$ in the act of, 
to put it graphically, being swept through by the object. 
Observe that this is not the ZPF of $I_\tau$ that in $I_\tau$ should be 
homogeneous and isotropic. 
For this we fix our attention on a fixed point 
of $I_*$, say the point of the observer at $(c^2/a, 0, 0)$ of $I_*$, 
that momentarily coincides with the object at the object initial proper time 
$\tau = 0$, and consider that point as referred to another inertial frame  
$I_\tau$ that instantaneously will coincide with the object at a future 
generalized object proper time $\tau > 0$. Hence we compute the  $I_\tau$-frame 
Poynting vector, but as instantaneously evaluated at the $(c^2/a, 0, 0)$ space point of the $I_*$
inertial frame, namely  in  $I_\tau$ at the  $I_\tau$ space-time point:

$$
ct_\t = - {c^2 \over a} \sinh \atc \ ,\ x_\t = - {c^2 \over a} \cosh \atc \ , \ y_\t=0 \ , \ z_\t
= 0 \ ,
\eqno(16)$$

\noindent
where the time in $I_\tau$, called $t_\tau$, is set to zero at the instant 
when $S$ and $I_\tau$ (locally) coincide, which happens at proper time 
$\tau$. (Here we correct 
a typo in Ref. [3], Eq.~(20) where the minus sign in the RHS of 
the $ct_\tau$ equation in eqn. (16) does not appear.) 
Everything, however, 
is ultimately referred to the $I_*$ inertial frame or laboratory frame. 
For further light on this point, see Appendix C of Ref. [3].
We first compute
the ZPF Poynting vector that enters the body of the accelerating object
in the instantaneous comoving frame $I_\tau$, 

$$\eqalignno {
{\bf S}_*^{zp} & = {c \over 4 \pi} \ZL \ {\bf E}_\t^{zp} \times {\bf B}_\t^{zp} \ \ZR_* 
& \cr
& = {c \over 4\pi} \Big\lbrace
\xh \ZL E_y B_z-E_zB_y \ZR + \yh \ZL E_z B_x - E_x B_z \ZR + \zh \ZL E_x B_y -
E_y B_x \ZR \Big\rbrace . 
& (17)} $$

\noindent
The star in the equation above implies that the quantity needs to be 
evaluated in the laboratory inertial frame $I_\star$. 
Since it turns out that only the two terms of the $x$-component of the ZPF Poynting 
vector are non-vanishing and the other 
seven components are zero, only these two non-vanishing terms will be 
examined here. For detailed calculations of all quantum averages, 
see Ref. [10]. Their values 
exactly match those of the SED analyses in Appendix A of Ref. [3].

In order to evaluate the vacuum expectation value $\ZL E_yB_z \ZR$, 
the $y$-component of
the electric field operators (13) and the 
$z$-component of the magnetic field operators (14), 
are multiplied together.  
The resulting expression has four terms, but as stated earlier only the term 
proportional to $\ZL \a \kl \ad\klp \ZR$
remains as in (5a), and the expression simplifies to

$$
\ZL E_y B_z \ZR = \S \int d^3k \Hzpsq
\Big[ \cosh \atc \eh_y - \sinh (\kh \times \eh)_z \Big]
\Big[ \cosh \atc (\kh \times \eh)_z - \sinh \atc \eh_y \Big]
\eqno(18)$$

\noindent
after one integration over the $k$-sphere.
Each of the four terms in the equation above may be evaluated using the 
following polarization equations,

$$\eqalignno {
\S \eh_y (\kh \times \eh)_z & = \kh_x \ , & (19a) \cr
\S (\kh \times \eh)_z (\kh \times \eh)_z & = \kh^2_x + \kh^2_y = 1 - \kh^2_z \ , & (19b) \cr
\S \eh_y^2 & = 1 - \kh_y^2 \ , & (19c) \cr
}$$

\noindent
and the expression becomes

$$
\ZL E_y B_z \ZR = \S \int d^3k \Hzpsq
\Big\lbrace \Big[
\cosh^2 \atc + \sinh^2 \atc \Big]\kh_x - \cosh \atc \sinh \atc
\Big[ 1 + \kh_x^2
\Big] \Big\rbrace \ . \eqno(20)
$$

\noindent
Compare this to Eq. (A30) in Ref. [3].
The first term above is zero since 
$\int d^3k~\hat{k}_x=0$. 
With the relation 
$\sinh\theta \cosh\theta = {1 \over 2} \sinh(2\theta)$
and the change of variable from $k$ to $\omega$, the expression
simplifies to 

$$
\ZL E_y B_z \ZR = - {4 \pi \over 3} \sinh \aatc \int \C.
\eqno(21) $$

The other non-vanishing term $\ZL E_zB_y \ZR$ can also be evaluated 
following the same procedure as above and it is found that these
two terms have the same magnitude but the opposite sign.
With these results, the Poynting vector ${\bf S}_*^{zp}$ of (17)
becomes

$$
{\bf S}_*^{zp} = - \xh {c \over 4 \pi} {8 \pi \over 3} \sinh \aatc \int \C .
\eqno(22) $$

\noindent
This represents the energy flux, i.e., the ZPF energy that enters the
uniformly accelerating object's body per unit area per unit time 
from the viewpoint of the observer at rest in the inertial 
laboratory frame $I_*$. The radiation entering the body is that of the 
ZPF centered at the $I_*$ frame of the observer. This leads to some 
detailed consideration on the so called $k$-sphere associated with each 
one of the inertial frames. For subtleties on this fine but important 
point, we refer to the Appendix C of Ref. [3].

The net impulse given by the field to the accelerated object, i.e., the 
total amount of momentum of the ZPF background the object has swept 
through after a time duration $t_*$, as judged again from the 
$I_*$-frame viewpoint, is 

$$
{\bf p}_* = {\bf g}_* V_* = {{\bf S}_* \over c^2} V_* = - \xh {1 \over c^2} {c \over 4 \pi}
\gt^2\beta_\t {2 \over 3} \big\langle{\bf E}_*^2 + {\bf B}_*^2 \big\rangle V_*.
\eqno(23) $$

\noindent
Combining this with Eq.(11), (12) and (22) we obtain 

$$
{\bf S}_*(\t) = \xh {c \over 4 \pi} \big\langle E_y B_z - E_z B_y \big\rangle =
\xh {c \over 4 \pi} {8 \pi \over 3} \sinh \aatc \int \C ,
\eqno(24) $$

\noindent
where as in Eqs. (21--23) the integration is 
understood to proceed over the $k$-sphere of $I_*$.
This $k$-sphere is a subtler 
point referring to the need to regularize certain {\it prima facie} improper 
integrals. ${\bf S}_*(\tau)$ represents energy flux, and it also implies 
a parallel, $x$-directed 
momentum density, i.e., field momentum growth per unit time and per unit 
volume as it is incoming towards the object position, $(c^2/a, 0, 0)$ of 
$S$, at object proper time $\tau$ and as estimated from the viewpoint of $I_*$. 
Explicitly such momentum density is

$$
{\bf g}_*^{zp}(\t) = {{\bf S}_*(\t) \over c^2} = 
- \xh {8 \pi \over 3} {1 \over 4 \pi c} \sinh \aatc \int \eta (\w) \C ,
\eqno(25) $$

\noindent
where we now introduce the henceforth frequency-dependent coupling or 
interaction coefficient $0 \leq \eta(\omega) \leq 1$, that quantifies 
the fractional amount of interaction at each frequency.

Let $V_0$ be the proper volume of the object. 
From the viewpoint of $I_*$, however, because of Lorentz contraction 
such volume is then $V_* = V_0/\gamma_\tau$. The amount of 
momentum due to the field inside the volume of the object according to 
$I_*$, i.e., the field momentum in the volume of the object 
viewed at the laboratory is

$$
{\bf p}_*(\t) = V_* {\bf g}_*= {V_0 \over \gt} {\bf g}_*(\t) = - \xh {4 V_0 \over 3} c \beta_\t \gt \Big( {1
\over c^2} \int \eta(\w) \C \Big) \ , 
\eqno(26) $$

\noindent
which is again Eq. (23).

At proper time $\tau = 0$, the $(c^2/a, 0, 0)$ point of the laboratory inertial 
system $I_*$ instantaneously coincides and comoves with the object point of 
the Rindler frame $S$ in which the object is fixed. The observer located 
at $x_* = c^2/a,~y_* = 0,~z_* = 0$ instantaneously, at $t_* = 0$, coincides and 
comoves with the object but because the latter is accelerated with constant 
proper acceleration ${\bf a}$, the object according to $I_*$ should 
receive a time rate of change of incoming ZPF momentum of the form:

$$
{d{\bf p}_* \over dt_*} = {1 \over \gt} {d{\bf p}_* \over d\t } \Big|_{\t=0} 
\eqno(27) $$

We identify this expression with a force from the ZPF on the object. 
If the object has a proper volume $V_0$, the force exerted on the object by 
the radiation from the ZPF as seen in $I_*$ at $t_* = 0$ is then

$$
{\bf f}_* = {d{\bf p}_* \over dt_*} = - \Big({4 \over 3}{V_0 \over c^2} \int \eta (\w) \C \Big) \ {\bf a} .
\eqno(28)$$

\noindent
Furthermore

$$
m_i = {V_0 \over c^2} \int \eta (\w) \C \Big)
\eqno(29) $$

\noindent
is an invariant scalar with the dimension of mass. Observe that in 
Eq. (29) we have neglected a factor of $4/3$. 
A fully covariant analysis (See Appendix D of Ref. [3])
shows that it should be replaced by unity. We show this covariant 
analysis in QED formulation in our Appendix B. The 
corresponding form of  
$m_i$ is then the mass of that 
fraction of the energy of the ZPF radiation enclosed within the object that interacts 
with the object as parametrized by the $\eta(\omega)$ factor in the integrand. 
Observe that $\eta(\omega)\rightarrow 0$ as $\omega \rightarrow \infty$  
because all bodies become transparent 
at sufficiently high frequencies. For further discussions on these 
developments we refer to the already published literature 
[3,12,13,16,17].

\vskip 10 pt
\noindent
{\bf 4. Relativistic four-force expression of Newton's Second Law}

\vskip 10 pt
\noindent
This analysis yields not just the nonrelativistic Newtonian case but also a fully 
relativistic description within special relativity, at least for the case of 
longitudinal forces, i.e., forces parallel to the direction of motion. 
Moreover the extension to the more general case, where the accelerating or 
applied force, ${\bf f}$, is non-uniform, (i.e., it changes both in magnitude and 
direction throughout the motion of the object), has been in principle 
accomplished [3].

From the definition of the momentum ${\bf p}_*$ in Eqs. (26, 29), 
it easily follows that the momentum of the body is
${\bf p}_* =m_i \gamma_\t \beta_\t c$,
in agreement with the momentum expression in special relativity. 
The space 3-vector component of the four-force [14] is then

$$
{\bf F}_* = \gt {d{\bf p_*}  \over  dt_* } ={d{\bf p_*}  \over  d\t } ,
\eqno(30) $$

\noindent
and as the force is pure in the sense of Rindler [14], 
the correct form for the four-force immediately follows,

$$
{\cal F}_* = {d{\cal P} \over d\t} =
{d \over d\t} (\gt m_i c, {\bf p} ) =
\gt \Big({1 \over c}{d E \over dt} , {\bf f} \Big) = 
\gt ({\bf f} \cdot {\bf \betaÐ_\t} , {\bf f} ) = 
({\bf F} \cdot {\bf \betaÐ_\t} , {\bf F} ) .
\eqno(31) $$

\noindent
Consistency with Special Relativity is established. A more detailed 
discussion leading to Eqs. (30)-(31)  appears in Ref. [3], 
in particular in its Appendix D.

We evaluated the Poynting vector of the ZPF radiation field that an
object under a constant proper acceleration (hyperbolic motion) sweeps through
as seen from the laboratory frame $I_*$, and found that there appears to 
exist an interaction between the object under hyperbolic motion and the 
ZPF (inertia reaction force), whose magnitude is proportional to the 
acceleration,
implying that
the ZPF possess a structure which reacts against acceleration.
We propose that this reaction force
between the accelerated object and the ZPF background
radiation is a part of what we know as inertia.

\vskip 10 pt
\noindent
{\bf Appendix A: Correspondence between SED and QED} 

\vskip 10 pt
\noindent
In this appendix, it is shown why the Poynting vector, 
${\bf S}={\displaystyle {c \over 4\pi}} ({\bf E}\times{\bf B})$,
indeed gives identical results for SED and QED averagings.
For this purpose, let us see the case of $\ZL E_yB_z \ZR$, one of the two 
non-vanishing terms.  
The other seven terms happen to vanish both in QED formulations [10] and 
in SED [3].

To evaluate $\ZL E_y B_z \ZR$, we multiply the $y$-component of ZPF 
electric field and the $z$-component of the magnetic field as given in 
Eqns.~(14,15) to obtain

$$\eqalignno {
\ZL E_y B_z \ZR & = \S \SP \int d^3 k \int d^3 k' 
{\sqrt {\hbar \w \over 2 \pi^2}} {\sqrt {\hbar \w' \over 2 \pi^2}}& \cr
& \times \cosh^2 \atc \Big[ \eh_y - \tanh \atc (\kh \times \eh )_z \Big] \Big[(\kh \times \eh)_z - \tanh \atc \eh_y
\Big]
\cr & \times {1 \over 2} \ZL
\big[ \a \kl e^{i\Theta} + \ad \kl e^{-i\Theta} \ \big]
\big[ \a \klp e^{i\Theta'} + \ad \klp e^{-i\Theta'} \ \big]
\ZR . & (32a) \cr 
}$$

\noindent
This equation has four terms. However, only the term proportional 
to $\ZL \a\kl \ad\klp\ZR$ remains non-vanishing due to
the vacuum expectation values given by Eqns. (5a, 5b).
Thus, after an intergration over the $k$-sphere, the right hand side of 
Eq. (32a) becomes 

$$
\ZL E_y B_z \ZR = {1 \over 2} \int d^3k {\hbar \w \over 2 \pi^2} \cosh^2 \atc \Big[\eh_y - \tanh \atc (\kh \times \eh)_z
\Big] \Big[ (\kh \times \eh )_z - \tanh \atc \eh_y \Big] .
\eqno(32b) $$

\noindent
This expression can be evaluated with the help of the polarization 
relations, Eq. (19), 
and the angular integration 
$\int \hat{k}^2_x d\Omega=\int \sin^3\theta d\theta \int \cos^2\phi d\phi=4\pi/3$, 
and we find that

$$
\ZL E_y B_z \ZR = - {4\pi \over c} \sinh \aatc \int \C ,
\eqno(33) $$

\noindent
which is the same value as the SED analogue $\left< E_y B_z \right>$, as 
already reported in Appendix A of Ref. [3].
It can be shown easily, following the same procedures, that the other 
non-vanishing term $\ZL E_z B_y \ZR$ also yields exactly the same 
value as the SED case. Thus, the correpondence between SED and QED is 
achieved, and 
we can indeed write 

$$
\langle {\bf S}\rangle = {c \over 4\pi} \langle \ {\bf E} \times {\bf B} \rangle = {c \over 4\pi}
\ZL \ {\bf\overline E} \times {\bf\overline B} \ \ZR = \ZL {\bf S} \ZR
\eqno(34) $$

\vskip 10 pt
\noindent
{\bf Appendix B: Covariant Approach}

\vskip 10 pt
\noindent
In this section, the electromagnetic ZPF Poynting vector 
${\bf S}^{zp}=\displaystyle{c \over 4\pi}\left({\bf E}^{zp}\times{\bf B}^{zp}\right)$
and its vacuum expectation values
are to be evaluated using a covariant method.
It will be shown, following the approach by Rohrlich [15], and 
Appendix D of Ref. [3], that the factor of $4/3$ for an 
expression of inertial mass, obtained earlier in the non-covariant 
method, vanishes in this fully covariant approach. 

The Poynting vector ${\bf S}$ is an element of the
symmetrical {\it electromagnetic energy-momentum tensor}

$$
\Theta^{\mu \nu}=
\pmatrix{-U & -S_x/c & -S_y/c & -SÐz/c \cr
-S_x/c & T_{xx} & T_{xy} & T_{xz} \cr
-S_y/c & T_{yx} & T_{yy} & T_{yz} \cr
-S_z/c & T_{zx} & T_{zy} & T_{zz}
} \eqno(35)
$$

\noindent
In the above, the time and mixed space-time components are

$$
\Theta^{00} = {1 \over 8\pi} (E^2 + B^2) \equiv -U ' \ {\rm and} \ \Theta^{0i} =
 - {1 \over 4\pi} ( {\bf E} \times {\bf B} )_i
\eqno(36) $$

\noindent
where $U$ is the electromagnetic energy density and
$S$ is the {\it Poynting vector}.
The space part of the tensor $\Theta^{ij}$ is the {\it Maxwell
stress tensor} whose components are given as

$$T_{ij} = {1 \over 4\pi} \Big[E_i E_j + B_i B_j - {1 \over 2} (E^2 + B^2) \delta_{ij} \Big]
\eqno(37) $$

Now let us consider the quantity,

$$
P^{\mu} \equiv {1 \over c} \int \Theta d\sigma_{\nu} =  \Big( {1 \over c} W, {\bf P} \Big)
\eqno(38)$$

\noindent
the integration of the energy-momentum tensor over a spacelike
plane $\sigma$ given by the equation
$n^\mu x_\mu + c\tau = 0,$
where $n^\mu$ is the unit normal vector to the three dimensional 
hyperplane.
Any instant of an inertial observer is characterized by this spacelike
plane $\sigma$ and the unit normal $n^\mu$. For example, when
$n^\mu=(1;0,0,0)$, $\tau=t$, then the spacelike plane $\sigma$
describes the $xyz$-plane at the instant $t$. For further details on 
this point, we refer the reader to Ref. [15] and 
Appendix D of Ref. [3].

In the particular Lorentz frame whose surface normal is given by 
$n^\nu=(1;0,0,0)$, the components of $P^\mu$ can be given explicitly as

$$
W^{(0)} = \int U^{(0)} d^3x \ , \ {\rm and,} \ {\bf P}^{(0)} = {1 \over c^2} \int {\bf S}^{(0)} d^3x .
\eqno(39)$$

However, in the case of interest to us in which the velocity is along the
positive $x$-direction, the surface normal is given by
$n^\nu=(\gamma;\gamma\beta {\bf {\hat n}})$, 
and Eq.~(38) takes the following forms: 

$$
W = \g \int U d\sigma - {\g\beta \over c} \int {\bf S \cdot \hat{\bf n}} d\sigma , \ {\rm and,}
 \ {\bf P} = {\g \over c^2} \int {\bf S} d\sigma + {\g\beta \over c} \int \Tarr
\cdot \hat{\bf n} d\sigma.
\eqno(40)
$$

At this point, we identify $P^\mu$ of Eq. (38) as the
momentum four-vector of the electromagnetic field. Note in passing that
extra terms appear in (40), which can also be
obtained from the corresponding Lorentz transformation. Abraham and Lorentz used the Eqns. (39)) 
as their definitions for the energy density and the momentum in the case 
of the Coulomb self-field of the classical electron, and they
were led to the incorrect factor of $4/3$ for the momentum of an 
electron. However, Eqns. (39) are only valid in
the particular Lorentz frame where $\gamma$ is 1, when the second terms 
in Eqns. (40) vanish.
We show that with the use of the correct forms Eqns. (40) for the
energy density and the momentum, this incorrect factor of $4/3$ is 
reduced to unity, as should be expected.

The expressions that we need to evaluate are 

$$
P^0 = {\g \over c} \int (U-{\bf v \cdot g} ) d^3\sigma, \ {\rm and ,}
\ {\bf p}_* = \g \Big( {\bf g}_* + {\Tarr_* \cdot {\bf v}_* \over c^2} \Big) V_0
\eqno(41) $$

\noindent
where the latter is the momentum 
of the background ZPF the object has swept through
as seen from the lab inertial frame $I_*$. 
The dot product of $\Tarr_*$ 
with the velocity ${\bf v}=v\xh$
in the above equation yields the column vector
$\Tarr_* \cdot {\bf v} = (\xh T_{xx} + \yh T_{yx} + \zh T_{zx} ) v $ 
with $T_{ij*}$ given by (37) and
$
\xh = \left[\matrix{1 \cr 0 \cr 0}\right] 
$, etc.  
It turns out that only the $x$-component has non-zero value, and the $y$ and $z$ 
components of the expectation values in 
$\Tarr \cdot {\bf v}$ vanish,
which is physically reasonable since the object is moving in the positive
$x$-direction. For the $x$-component, we have 

$$\eqalignno {
\ZL T_{xx*} \ZR & = {1 \over 4\pi} \ZL E_{x\t} E_{x\t} + B_{x\t} B_{x\t} - {1 \over 2} (E^2_\t + B^2_\t ) \ZR_* &
\cr  & = {1 \over 4\pi} \ZL E^2_{x\t} + B^2_{x\t} \ZR_* - {1 \over 8\pi} \ZL E^2_\t + B^2_\t \ZR_* & (42) \cr
}$$

\noindent
where

$$
E^2_\t = E^2_{x\t} + E^2_{y\t} + E^2_{z\t} , \ {\rm and} \
B^2_\t = B^2_{x\t} + B^2_{y\t} + B^2_{z\t} .
\eqno(43) $$

\noindent
The first term becomes

$$\eqalignno {
{1 \over 4\pi} \ZL E^2_{x\t} + B^2_{x\t} \ZR_* & = {1 \over 4\pi} \ZL E^2_{x*} + B^2_{x*} \ZR  \cr
& =  {1 \over 12\pi} \ZL E^2_* + B^2_* \ZR , & (44) \cr
}$$

\noindent
considering equal contributions from each direction. 
After the substitution of 

$$
U = {1 \over 8\pi} \ZL E^2_* + B^2_* \ZR = \int \C
\eqno(45) $$

\noindent
we find that for the first term of (42),

$$
{1 \over 4\pi} \ZL E^2_{x\t} + B^2_{x\t} \ZR_* = {2 \over 3} \int \C .
\eqno(46) $$

For the evaluation of the second term of (42), we find the 
Lorentz transformed field components from Eqns. (13,14),
and notice that the squared fields have contributions given by 
Eq. (45) to obtain

$$
\ZL T_{xx*} \ZR = {1 \over 3} \int \C (1 - 2\g^2_\t - 2 \g^2_\t \beta^2_\t ) .
\eqno(47) $$

\noindent
Using the two results above, we can obtain for the momentum

$$
{\bf p}_* = \gt \Big( {\bf g}_* + {\Tarr_* \cdot {\bf v}_* \over c^2} \Big) V_0 = \xh \gt V_0 c \beta_\t {1 \over
c^2} \int \C .
\eqno(48) $$

\noindent
We note here that the extra factor of
$4/3$ obtained earlier in a non-covariant method becomes unity, as expected,
in this covariant approach.

Following similar steps, we can also evaluate the zero-component of the 
momentum four-vector as

$$
P^0 = {\gt \over c} \Big[{\ZL E^2_* + B^2_* \ZR \over 8\pi} - c\beta_\t g_* \Big] V_0 + {\gt V_0 \over c} \int \C .
\eqno(49) $$

The inertia reaction force that is exerted upon the object by the ZPF is

$$\eqalignno {
{\bf f}_*^{zp} & = - {d{\bf p}_* \over dt_*} = - {1 \over \gt} {d{\bf p} \over dt_*} & \cr
& = - \Big( {V_0 \over c^2} \int \eta(\w) \C \Big) \ {\bf a} & (50) \cr 
}$$

With the identification of the quantity inside the parenthesis
as the inertial mass $m_i$, we can obtain 
the four-force as

$$
F^\mu = {dP^\mu \over d\t } = {d \over d\t}  (m_i c \gt ; {\bf p}) 
= \gt \Big( {1 \over c} {dE \over dt} ; {d{\bf p} \over dt} \Big)
= \gt ( {\bf f} \cdot \beta_\t ; {\bf f} )
= \gt ( {\bf F} \cdot \beta_\t ; {\bf F} )
\eqno(51) $$

\noindent
which is the same expression as the Eq. (31) above.

\vskip 10 pt
\noindent
{\bf References}

\vskip 10 pt
\parskip=0pt plus 2pt minus 1pt\leftskip=0.25in\parindent=-.25in 
  
[1] T. S. Kuhn, {\it Black-Body Theory and the Quantum Discontinuity, 1894-1912},
        {Clarendon, Oxford}, {1978}. This includes a 
	thorough and detailed account of the discovery of the zero-point 
	energy by Planck (ca.~1912) and rapid appreciation of this concept by 
	Einstein and Stern (ca.~1913), and later by others.

[2] B. Haisch, A. Rueda and H.~E. Puthoff, {\it Phys. Rev. A} 
	        {\bf 49}, 678 (1994).

[3] A. Rueda and B. Haisch, {\it Found. Phys.} {\bf 28}, 1057 (1998). In this paper it is argued why a more
satisfactory approach is needed to replace that of [2]. A detailed discussion of this comparison is deferred for a
future paper. See also A. Rueda, B. Haisch and Y. Dobyns, {\it Annalen der Physik} (Leipzig), {\bf 10}, 393 (2001).
http://arxiv.org/abs/gr-qc/0009036

[4] W. P. Healy, {\it Non-Relativistic Quantum Electrodynamics}, Academic Press, London (1982).

[5] T.~H. Boyer, {\it Phys. Rev.} {\bf 182}, 1374 (1969). 

[6] M. Planck, {\it Annalen der Physik} {\bf 37}, 642 (1912).

[7] A. Einstein and L. Hopf, {\it Annalen der Physik} {\bf 33}, 1105, (1910).

[8] T.~H. Boyer, {\it Phys. Rev. D} {\bf 11}, 809 (1975). 

[9] W. H. Louisell, {\it Quantum Statistical Properties of Radiation},
        {John Wiley \& Sons, Inc. New York}, {1990}.

[10] H. Sunahata, {\it Interaction of the quantum vacuum with an accelerated 
                 object and its contribution to inertia reaction force}, Ph.D. thesis, 
                 Claremont Graduate University, (CA, USA, 2006).\break
                 http://www.calphysics.org/articles/QED\string_Inertia.pdf

[11] L.~de la Pe$\tilde{\rm n}$a and A.~M. Cetto, {\it The Quantum Dice---An Introduction to Stochastic
Electrodynamics}, Kluwer Academic Publishers, Dordrecht, (1996).

[12] A. Rueda and B. Haisch, {\it Phys. Lett. A} {\bf 240}, 115 (1998). http://arxiv.org/abs/physics/9802031

[13] A. Rueda and B. Haisch, {\it Annalen der Physik} (Leipzig) {\bf 14} (8), 
               479 (2005). http://arxiv.org/abs/gr-qc/0504061

[14] W. Rindler, {\it Introduction to Special Relativity}, Clarendon, Oxford, (1991).

[15] F. Rohrlich, {\it Classical Charged Particles} Addison-Wesley, Reading, (1965).

[16] M. Jammer, {\it Concepts of Mass} Princeton University Press, Princeton, NJ, (2000). 
                 See, in particular, pp. 163-167.

[17] Y. Dobyns, A. Rueda and B. Haisch, {\it Found. Phys.} {\bf 30}, 59 (2000). http://arxiv.org/abs/gr-qc/0002069

\bye